\documentclass{webofc}

\usepackage[varg]{txfonts}   
\usepackage{hyperref}
\usepackage{url}
\usepackage{amsmath}
\usepackage{bm}
\hypersetup{colorlinks=true,citecolor=blue,urlcolor=blue,linkcolor=blue}

\begin{document}

\title{Hydrodynamic Short-Range Correlations from Boltzmann-Langevin Equation}

\author{Li Yan\inst{1}\fnsep\thanks{\email{cliyan@fudan.edu.cn}} \and
        Derek Teaney\inst{2}
}

\institute{Institute of Modern Physics, Fudan University, Shanghai, China
\and
           Department of Physics and Astronomy, Stony Brook University, New York, USA
}

\abstract{We investigate hydrodynamic contributions to short-range two-particle correlations in relativistic heavy-ion collisions using the Boltzmann-Langevin equation. 
We derive and solve the transport equation for equal-time two-point correlations, obtaining both local and non-local contributions that scale with transport coefficients. The non-local correlations emerging from 2-to-2 scattering dynamics provide a hydrodynamic signature in short-range correlation measurements.}

\maketitle

\section{Introduction}
\label{intro}

In relativistic heavy-ion collisions, two-particle correlations have been extensively studied as probes of quark-gluon plasma dynamics. While long-range (in pseudo-rapidity) correlations are well-established signatures of collective flow~\cite{Shen:2020mgh}, from which harmonic flow of different orders has been quantitatively characterized via a hydrodynamical modeling the system evolution~\cite{Heinz:2013th}. On the other hand, the short-range correlations that have been predominantly attributed to "non-flow" effects such as those from jet fragmentation, are substracted in general. However, owing to the therrmal fluctuation inside the medium of quark-gluon plasma, fluctuations of hydrodynamical variables, such as temperature $\delta T$ and flow velocity $\delta u^\mu$, are essentially correlated from space and time. Accordingly, the emitted particles from the medium, can in principle be correlated, irrespective of their separation of kinematic range. In paricular, short-range correlations of two particles receive contributions from the late-time thermal fluctuations which are consequences of collective flow, containing transport properties of the medium.

This work demonstrates that hydrodynamic fluctuations~\cite{Kapusta:2011gt} (thermal fluctuations in fluids), can in principle generate significant short-range correlations through the conversion of thermal fluctuations in the fluid to correlated particle emissions. 
While we do not solve explicitly the stochastic hydrodynamic evolution, we focus on the resulting two-particle correlation which has a deterministic solution. By employing the Boltzmann-Langevin framework to systematically compute the two-particle correlations, fluctuating hydrodynamics can be related to the observed particle correlations through the freeze-out process.

\section{Boltzmann-Langevin Framework}
\label{formalism}


The conversion of fluid fluctuations to particle correlations occurs during freeze-out. For a fluctuating fluid described by thermodynamic variables $(\alpha = \mu/T, \beta = 1/T, u^\mu, e, P)$ with fluctuations $(\delta\alpha, \delta\beta, \delta u^\mu, \ldots)$, the two-particle correlation becomes:

\begin{equation}
\{\delta N_p\delta N_q\} \propto \{\delta f_p\delta f_q\} \neq 0,
\end{equation}
with $\delta f_p \equiv f_p - \{ f_p\}$, and hence in the deterministic case one expects $\{\delta f_p\}\{\delta f_q\} = 0$. These correlations are governed by fluctuation-dissipation relations and scale with transport coefficients.

The Boltzmann-Langevin equation provides a stochastic description~\cite{FOX,Calzetta:1999xh}:
\begin{equation}
\frac{df}{dt} = [\partial_t + \bm{v} \cdot \nabla]f = -\mathcal{C}[f] + \xi_p(\bm{x},t),
\label{eq:bl}
\end{equation}
with noise correlation with respect to the fluctuation-dissipation relation,
\begin{equation}
\{\xi_p(\bm{x},t)\xi_{p'}(\bm{x}',t')\} = 2\mathcal{K}_{pp'}\,\delta(\bm{x}-\bm{x}')\delta(t-t').
\label{eq:noise}
\end{equation}
Note that the linearized collision kernel, $\mathcal{K}_{pp'}$, gives rise to the transport coefficients after integration over the momentum space~\cite{Arnold:2000dr}. For instance, in the relaxation time approximation (RTA), $\mathcal{K}_{pp'}=n_p' \delta_{pp'}/\tau_R$, in which $\eta/Ts \propto \tau_R$.
The consistency of the Boltzmann-Langevin equation can be examined as well through its relation to fluctuating hydrodynamics, and one indeed has the identifications:
\begin{equation}
T^{\mu\nu} = \int_p \frac{P^\mu P^\nu}{E_p} f_p, \quad S^{\mu\nu} = \int_p \frac{P^\mu P^\nu}{E_p} \mathcal{C}_p^{-1} [\xi_p],
\end{equation}
which yield the expected fluctuation-dissipation relation~\cite{landau1980statistical},
$
\{S^{\mu\nu}S^{\alpha\beta}\} = 2T\eta\Delta^{\mu\nu\alpha\beta}\delta_{xx'}.
$


The central object of our study is the equal-time two-point correlation function:

\begin{equation}
\Lambda_{p_1 p_2}(\bm{x}_1, \bm{x}_2, t) \equiv \{\delta f_{p_1}(\bm{x}_1, t)\delta f_{p_2}(\bm{x}_2, t)\},
\label{eq:correlation}
\end{equation}
which directly relates to observable particle correlations:
\begin{equation}
\{\delta N_p \delta N_q\} = \int_{\Sigma} d\sigma_{1\mu} d\sigma_{2\nu} p^\mu q^\nu \Lambda_{pq}.
\end{equation}
From the Boltzmann-Langevin equation, we derive the deterministic transport equation for the two-point correlator:
\begin{equation}
[\partial_t + \bm{v}_1 \cdot \nabla_1 + \bm{v}_2 \cdot \nabla_2]\Lambda_{p_1 p_2} = -\left(\mathcal{C}_{p_1}[(n'_{p_1})^{-1}\Lambda_{p_1 p_2}] + \mathcal{C}_{p_2}[(n'_{p_2})^{-1}\Lambda_{p_1 p_2}]\right) + 2\mathcal{K}_{p_1 p_2}\delta(\bm{x}_1 - \bm{x}_2).
\label{eq:transport}
\end{equation}

The local equilibrium solution of the equation can be proved to be,
\begin{equation}
\Lambda^{\text{eq}}_{p_1 p_2}(\bm{x}_1, \bm{x}_2, t) = (2\pi)^3 n'_{p_1} \delta_{p_1 p_2} \delta(\bm{x}_1 - \bm{x}_2),
\label{eq:equilibrium}
\end{equation}
which serves as the zero mode of the collision operator. This form can be derived from entropy considerations:
\begin{equation}
\delta s = -\frac{1}{2} \int\frac{ d^3 \vec p}{(2\pi)^3} \frac{(\delta f_p)^2}{n'_p}\,.
\end{equation}
Eq.~(\ref{eq:equilibrium}) is consistent with the well-known equilibrium correlations among therrmal fluctuations. For instance, one has
$
\{\delta e \delta e\}_{\text{eq}} = \frac{T^2 s}{c_s^2} \delta_{xx'}$ and $\{\delta n \delta n\}_{\text{eq}} = T\chi_n \delta_{xx'}$,
where $c_s^2$ and $\chi_n$ are the square of speed of sound and the charge susceptibility, respectively. Below, we solve the leading order correction to the two-point correlator from dissipation~\cite{Teaney:2013gca}, in the context of the RTA and the scalar theory. 

\begin{enumerate}
\item[I.] Relaxation Time Approximation
\end{enumerate}

In RTA, $\mathcal{K}_{pp'} = (n'_p \delta_{pp'}) / \tau_R$, we obtain the analytical solution:
\begin{equation}
\Lambda^{(1)}_{p_1 p_2} = -\frac{\tau_R}{2} \frac{n'_p}{TE_p}(\chi^{\text{dif}}_{p} + \chi^{\text{vis}}_{p})\delta_{p_1 p_2}\delta(\bm{y}),
\label{eq:rta-sol}
\end{equation}
where $\bm{y}=\bm{x}_1-\bm{x}_2$ denotes the separation in space, and the scalar response functions are:
\begin{equation}
\chi_p^{\text{diff}} = -TP^\mu\nabla_\mu\alpha, \quad \chi_p^{\text{vis}} = P^\mu P^\nu\langle\nabla_\mu U_\nu\rangle.
\end{equation}
This solution is proportional to transport coefficients ($\eta/s$, $D/T^2$) but is purely local in both momentum and coordinate space.
\begin{enumerate}
\item[II.] 2-to-2 Scatterings in Scalar Theory
\end{enumerate}
For a more realistic $\lambda\phi^4$ theory with transport coefficients,
$
\eta = C_\eta \frac{T^3}{\lambda^2}$ and $\sigma = C_\sigma \frac{T}{\lambda^2},
$
where $C_\eta$ and $C_\sigma$ are constants to be determined, 
the collision kernel decomposes into mean-field and s-, t- and u-channel contributions:
\begin{equation}
\frac{dn'_{p_1}}{dt} \delta_{p_1 p_2} \delta(\bm{y}) = -\sum \left\{ \mathcal{I}_{p_1} \Phi^{(1)}_{p_1 p_2} + \int_{p'} \mathcal{I}^{(s)}_{p_1 p'} \Phi^{(1)}_{p' p_2} - \int_{k} \mathcal{I}^{(t)}_{p_1 k} \Phi^{(1)}_{k p_2} - \int_{k'} \mathcal{I}^{(u)}_{p_1 k'} \Phi^{(1)}_{k' p_2} + (p_1 \leftrightarrow p_2) \right\}.
\end{equation}
We decompose the solution into singular (local) and regular (non-local) parts:
\begin{equation}
\Phi^{(1)}_{p_1 p_2} = \rho^{(s)}_{p_1} \delta_{p_1 p_2} \delta(\bm{y}) + \rho^{(r)}_{p_1 p_2} \delta(\bm{y}).
\label{eq:decomposition}
\end{equation}
The singular part is solved analytically:
\begin{equation}
\rho^{(s)}_{p} = -\frac{1}{2\mathcal{I}_{p}} \frac{dn'_{p}}{dt} = -\frac{(1+2n_{p})n'_{p}}{2TE_{p}\mathcal{I}_{p}}[\chi^{\text{dif}}_{p} + \chi^{\text{vis}}_{p}],
\end{equation}
while the regular part satisfies an integral equation:
\begin{equation}
\mathcal{I}_{p_1} \rho^{(r)}_{p_1 p_2} + \int_k G_{p_1 k} \rho^{(r)}_{k p_2} + (p_1 \leftrightarrow p_2) = S_{p_1 p_2},
\end{equation}
with effective Green's function and source:
\begin{equation}
G_{pq} = \mathcal{I}^{(s)}_{pq} - \mathcal{I}^{(t)}_{pq} - \mathcal{I}^{(u)}_{pq}, \quad S_{pq} = -(\rho^{(s)}_{p} G_{qp} + \rho^{(s)}_{q} G_{pq}).
\end{equation}

\begin{figure}[h]
\centering
\includegraphics[width=5cm]{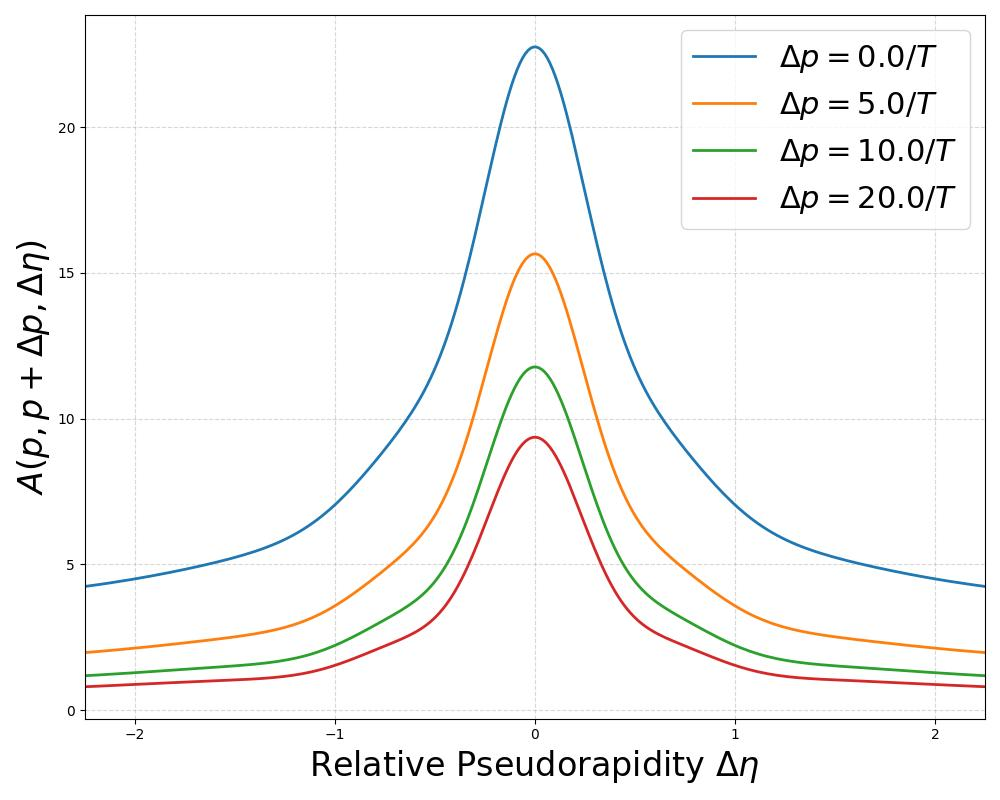}
\includegraphics[width=5cm]{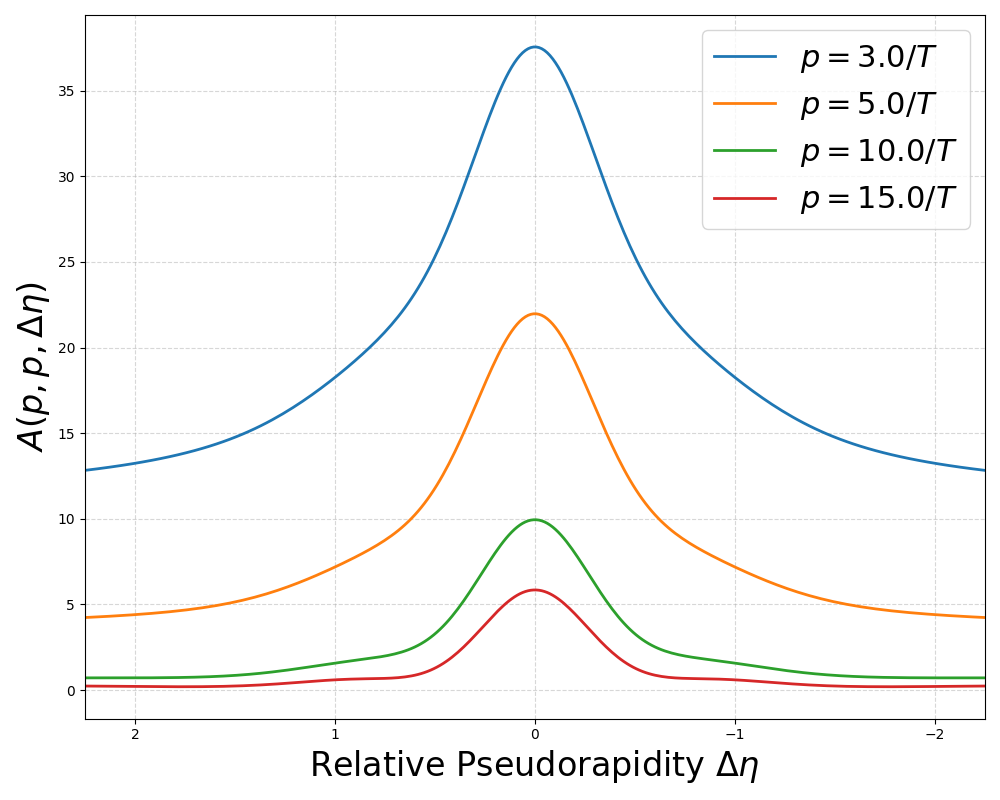}
\caption{Numerical solution of the two-point correlation function, with the relative angle $\theta_{pq}$ chosen with respect to the effective gap of pseudo-rapidity $\Delta\eta$.}
\label{fig:Afunc}
\end{figure}

The regular correlation function admits a formal solution:
\begin{equation}
n_{p_1}\rho^{(r)}_{p_1 p_2} = \lambda^{-2} \left[A(p_1, p_2, \theta_{12})\chi_{p_2} + A(p_2, p_1, \theta_{21})\chi_{p_1}\right],
\end{equation}
where $A(p,q,\theta_{pq})$ can be solved numerically via Legendre expansion:
\begin{equation}
A(p,q,\theta_{pq}) = \sum_{n=0}^\infty A_n(p,q) P_n(\cos\theta_{pq}).
\end{equation}
In Fig.~\ref{fig:Afunc}, we present the numerical results of the two-point correlation function in terms of $A(p,q,\theta_{pq})$. For the purpose of illustration, we choose and convert the relative angle $\theta_{pq}$ to the gap of pseudo-rapidity $\Delta\eta$, to mimic the emission of two particles in heavy-ion collisions on top of a Bjorken flow background. As can be seen from the figure,  we find non-trivial angular structure in momentum space correlations, which exhibits thermal suppression effects at higher momenta. These non-local correlations provide a hydrodynamic signature in short-range correlation measurements that was previously attributed entirely to non-flow effects.

\section{Conclusion}
\label{conclusion}

We have demonstrated that short-range two-particle correlations in heavy-ion collisions contain significant hydrodynamic contributions from late-time collective dynamics. Using the Boltzmann-Langevin framework, we derived the complete structure of the two-point correlation function, up to the leading order dissipative corrections, in the RTA and the scalar theory.
%
The non-local correlations emerging from 2-to-2 scattering dynamics provide a unique hydrodynamic signature that should be observable in high-precision correlation measurements at RHIC and LHC.


\section*{Acknowledgements}
LY is supported in part by the National Natural Science Foundation of China through Grants No. 12375133 and No. 12147101.

%
%
%
%
%

\bibliographystyle{abbrv}
\bibliography{refs-qm25}

\end{document}